# High-Throughput, Semi-Autonomous Measurement of Cavitation-Mediated Material Breakage

David G. Bell,[1] Matthew A. Hopcroft,[1] and William M. Behnke-Parks[1]

[1]*Applaud Medical, 953 Indiana St., 94107, San Francisco, California, USA*

Engineered microbubbles can be acoustically driven to cavitate against a substrate to produce erosion and fragmentation. This mechanical action has therapeutic applications in the treatment of biomineralizations, such as in urinary stone disease. However, current methods for quantifying the mechanical action of cavitation on a substrate are slow or imprecise. In this paper, we describe the design of a device that applies calibrated pressures to microbubbles engineered to target a calcium-containing hydroxyapatite substrate under physiological conditions and quantifies the result via an automated submerged mass measurement with high precision and low drift. Measurements of microbubble-mediated mass loss were observed to be linear with time, with variance that was comparable to the resolution of the instrument. The rate of mass loss with microbubbles present was 5.5-fold greater than in the absence of microbubbles. The research instrument described here captures the essential mechanical and physiological features of *in vivo* microbubble-mediated erosion and fragmentation of urinary stones and has been used to optimize the parameters of this treatment modality for a clinical trial of this promising new therapy for nephrolithiasis. In addition to clinically relevant therapeutic applications, this approach will contribute to broader understanding of acoustic cavitation against a substrate.

## I. INTRODUCTION

The mechanical action of engineered microbubbles (diameters of 0.1-10 μm) underpins a number of therapeutic and diagnostic techniques, including cavitation-mediated erosion and fragmentation of biomineralizations and ultrasonic imaging.[1, 2] Incorporating components that have adherent properties into the shells of engineered microbubbles enhances many of these applications, enabling selective adherence to tissues and materials of interest to increase both efficacy and safety.[3] The use of targeted engineered microbubbles in the treatment of renal and urinary calculi provides a safe and effective approach to lithotripsy whereby cavitation focuses low intensity extracorporeal acoustic driving pressure to achieve higher local intensities.[2, 4] A treatment based on this approach has recently entered medical device clinical trials.

Given the promise of engineered microbubbles, there is substantial interest to better understand and optimize their mechanical action through quantitative methods. The quantitation of microbubble action on a target is a difficult problem: pressure focusing via inertial cavitation, convolved with the fracture mechanics of the material of the target, generates target breakage that is intrinsically non-linear, therefore requiring a careful and precise experimental approach. Prior experimental methods for characterizing calculus erosion and fragmentation, including imaging, profilometry, and endpoint techniques (e.g. fragment sieving, time-to-complete-comminution), are not conducive to continuous, quantitative data collection.[5] Here, we develop an instrument to apply calibrated pressures to engineered microbubbles localized to calculi and precisely measure the resulting loss of calculus surface material due to erosion and fragmentation. Termed submerged comminution analysis (SCA), this approach extrapolates true mass loss from changes in weight of submerged calculi, enabling rapid iteration through parameters relevant to optimizing the erosion process and understanding the mechanical action of microbubbles. Furthermore, SCA allows continuous data collection without handling of calculi or exposing calculi to air, which can induce artificial fragility.[5] Importantly, SCA also enables the accurate simulation of clinically relevant *in vivo* conditions. Calculi are placed within an artificial ureter oriented parallel to the ground to mimic a patient lying prone, and experiments can be run at elevated temperatures and in the presence of surrogate urinary flow.

Because change in calculus mass resulting from microbubble-mediated erosion can vary by orders of magnitude depending on the chosen set of parameters and the intrinsic mechanical strength of the calculus, high-resolution, low-drift measurement of mass loss is a salient feature of this work. The design principles presented here are applicable to a broader class of instruments, including overhanging radiation force balances for calibrated power measurements of ultrasonic transducers. We briefly discuss how features of this design can be applied to other instruments.

## II. EXPERIMENTAL APPARATUS

### A. Overview

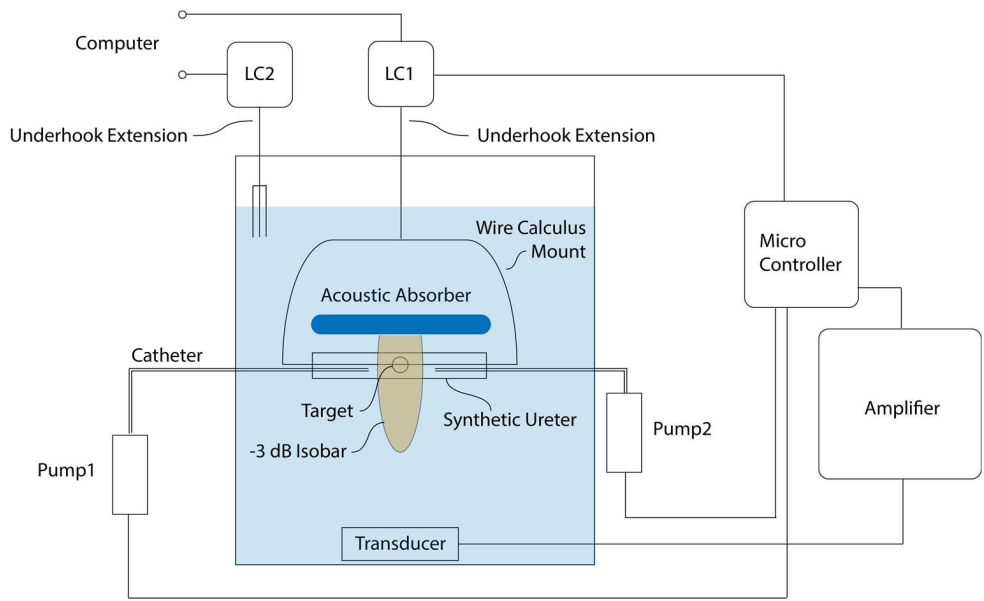

FIG. 1. Schematic overview of the apparatus, including fluidic handling of microbubbles (via Pump1) inside a synthetic ureter, precision low-drift load cells (LC1 and LC2), a quasi-collimated pressure transducer to produce a wide, uniform pressure distribution, and synchronized triggering via a microcontroller for high-throughput data collection.

The developed instrument (Fig. 1) comprised a load cell (LC1) coupled to an artificial calculus via a thin carbon fiber extension to the load-bearing underhook of LC1, wire calculi mount (affixed to the extension), and cable (used to suspend the stone between the two extremes of the calculi mount). In experiments, the stone was suspended inside an independently mounted model ureter (dialysis tubing) positioned via optical posts (Thorlabs). A transducer was mounted below the calculus and tubing such that its region of maximum pressure output aligned with the stone. An acoustic-absorbent pad was positioned between the model ureter and calculi mount to shield LC1 from excess acoustic energy, as well as prevent unwanted reflections off of the air-water interface. A second load cell (LC2) measured drift due to changes in water level by breaking the air-water interface with three underhook extensions identical to the one coupled to LC1. Microbubbles were delivered via a catheter attached to an external syringe pump (Pump1, Fig. 1). The pump was mounted to a modified nutator used to keep the distribution of bubbles within their suspending fluid homogenous. A second syringe pump (Pump2, Fig. 1) was used to withdraw tank fluid to offset infusion volumes. Further details regarding the setup will be discussed in the following sections.

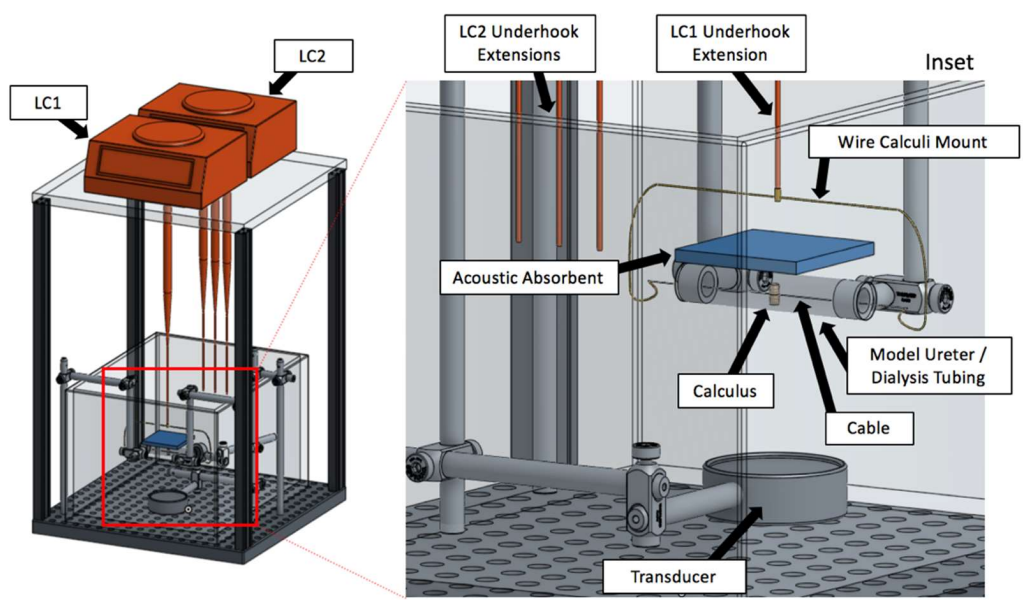

**FIG.** 2. Diagram of experimental setup: two load cells, LC1 and LC2, are positioned on a platform mounted above the tank and are coupled to underhook extensions that break the air-water interface. Inset: the underhook extension of LC1 is coupled to the wire calculi mount, cable, and stone, which is surrounded by dialysis tubing. The transducer is positioned below the stone, and an acoustic absorbent pad mitigates application of acoustic force on LC1. The three extensions coupled to LC2 break the air-water interface.

### B.  Microbubble infusion scheme

Engineered microbubbles incorporating a calcium-adherent component in a lipid shell (Applaud Medical) were used for erosion experiments presented here. Microbubbles suspended in Dulbecco's phosphate buffered saline (DPBS) will naturally stratify due to differences in buoyancy. To guarantee microbubble infusions of consistent size distribution and number density, the syringe pump used for microbubble delivery was mounted onto a modified nutator (HulaMixer® Sample Mixer, Thermo Fisher) capable of 360° rotation. The nutator's tube-holder plate was replaced with an aluminum strip to which the syringe pump was bolted. To avoid excessive twisting of the catheter coupled to the syringe pump, the nutator was programmed to reverse direction of rotation every 720°.

A syringe pump (NanoJet, Chemyx) was employed to deliver precise quantities of bubbles at consistent rates. The primary criteria for the selection of pump were weight, allowing the syringe and pump to be mounted on the nutator, and RS232 interface, permitting control via a governing circuit. Pump infusion rates and catheters were chosen such that they emulated their clinical counterparts and preserved bubble concentration. To this end, 5 Fr ureteral catheters (Open-End Ureteral Catheter, Cook Medical) were tested for bubble preservation and used in experiments. In the experiments presented here, engineered microbubbles were infused within 15 minutes of syringe placement onto the nutator.

## C. Measurement of mass

### 1. Coupling the calculus to the load cell

The coupling between calculus and LC1 (MS104TS, Mettler Toledo, 100 µg resolution) was designed to enable accurate mass measurement through the air-water interface, position the calculus within the model ureter, and mitigate the influence of radiation forces. The connection was realized as three components: a slim, homogenous rod to break the air-water interface, a wire calculi holder and cable to suspend the calculus within the model ureter, and an acoustic-absorbent pad to protect the LC1 from radiation forces.

*a. Breaking the air-water interface with a thin, homogenous rod*

A carbon fiber rod (3.05 mm diameter) was affixed to the underhook of LC1 such that the rod broke the air-water interface. The rod was chosen to minimize and linearize the buoyant and wetting forces ($F_b$ and $F_w$, respectively) on the scale, which may be expressed as:

$$F_b = \rho_L g A y \tag{1}$$

$$F_w = \gamma p \cos\theta \tag{2}$$

Here, $\rho_L$ is the density of the tank fluid, $g$ is the gravitational constant, $A$ is the cross-section area of the rod, $y$ is the immersion depth, $\gamma$ is the fluid surface tension, $p$ is the rod perimeter, and $\theta$ is the contact angle. A minimal diameter decreased the magnitude of both forces, while the smooth, homogenous surface and geometric uniformity of the carbon fiber mitigated contact angle hysteresis and pinning, which can lead to non-linearity in the surface tension force.[6, 7]

*b. Suspension of the calculus within model ureter*

A wire calculi holder, cable, and o-ring were used as intermediaries between the calculus and the underhook extension of LC1. The holder was mounted directly to the underhook extension, and the cable was affixed to each of the holder's extremes. A rubber o-ring was used to couple the cable and calculus, positioning the latter in the center of the model ureter. Importantly, the calculus, cable, and holder remained decoupled from the model ureter. To ensure the surface area of the calculus exposed to the transducer remained constant during experiments, the o-ring encircled the calculus at one of its extremes, preventing it from rotating about the axis of the cable as the calculus lost mass. The ureter was modeled with dialysis tubing (Carolina Biological Supply). In general, hydrophilic cellulose film was found to prevent adhesion of air, which could produce acoustic shielding and lower transmitted acoustic pressure. Pressure measurements via a needle hydrophone (HRN-0500, Onda Corp) found negligible change in pressure when the acoustic path was obstructed with dialysis tubing.

c. *Protecting LC1 from acoustic energy*

An acoustic-absorbent pad protected the calculi holder and LC1 underhook extension from the influence of acoustic radiation forces; solely the cable and calculus were left exposed. To further minimize the impact of such forces on measurement, the acoustic energy source was positioned directly below the calculus to avoid applying radiation force perpendicular to the axis of sensitivity of the scale.

d. *Effect of buoyancy on measurement*

Buoyant forces must be accounted for when weighing submerged calculi. Specifically, changes in calculus mass have accompanying changes in volume, which affect the buoyant force through Archimedes' principle. Accordingly, the weight change measured by LC1 is dictated both by the mass change of the calculus and by changes in the buoyant forces exerted on the calculus and underhook extension:

$$\delta m' g = \delta m g - \rho_L g \, (\delta V_c + A \, \delta y) \qquad (3)$$

Here, $\delta m'$ is the change in calculus mass LC1 perceives (load cells assume gravity is the only active downward force), $\delta m$ is the true change in calculus mass, $\rho_L$ is the density of the medium, $\delta V_c$ is the change in calculus volume, A is the cross-sectional area of the rod, and $\delta y$ is displacement of the LC1 underhook extension relative to the water line that results from a change in calculus mass. Under the assumptions that the calculus is of constant density and that LC1, an electromagnetic force compensation balance, does not deflect when loaded,[8] and assuming no other source of change in the air-water meniscus:

$$\delta V_c = \frac{\delta m}{\rho_c}, \qquad (4)$$

$$\delta y = 0 \qquad (5)$$

$\rho_c$ is the density of the calculus. From (2) and (3):

$$\delta m = \left(1 - \frac{\rho_L}{\rho_c}\right)^{-1} \delta m' \qquad (6)$$

Given that calculi employed here were denser than both water and urine, buoyant forces had non-negligible effect on measurement. Prior to experiments with a particular model of calculus, the average density of the strain was calculated by measuring both the dry and submerged weights of the stone. The density of the calculus was then computed as:

$$\rho_c = \rho_L \left(1 - \frac{\delta m\prime}{\delta m}\right)^{-1} \tag{7}$$

**D. Implementation of a second load cell for drift correction**

From (1), it is clear that fractional volume of underhook extension submerged can impact measured mass. This error in measurement is:

$$\delta m_{error} = \rho_L A\, \delta y \tag{8}$$

Here, $\delta y$ is due to change in the water line. In the described setup, this corresponds to a 7.3 mg error per millimeter change in water line for a carbon fiber rod with a 3.05 mm diameter. Pump infusions and evaporation both impact the water line and therefore contribute to this error. To counteract this effect, a second load cell, LC2 (Entris, Sartorius, 100μg resolution), was coupled to three additional identical carbon fiber rods that broke the air-water interface. As discussed in later sections, coupling three underhook extensions to LC2 and normalizing measured drift by the number of extensions reduced RMS noise when tracking the water line.

**E. Environmental Control**

To further limit the effects of evaporation, air convection, and temperature fluctuations on the apparatus, an enclosure encompassed the experimental setup. This enclosure was decoupled from the instrument itself to prevent external contact from influencing measurement and was outfitted with a door for convenience. The addition of the enclosure noticeably reduced LC1 settling time and noise as well as drift measured by LC2.

The enclosure additionally enabled experimentation at higher temperatures, which was critical for simulating *in vivo* conditions, where water temperature can impact bubble dynamics.[9] Prior to experiments, a convective heating element (Anova) was employed to elevate the tank water temperature. Because turbulence generated by the built-in propeller of the heating element produced noise in measurements, an identical heating element with propeller removed maintained tank temperature during experiments. The second non-convection heating element was less effective at maintaining a uniform temperature throughout the tank, but the strategy was still suitable for periods of ~30 minutes. For temperatures above 30C, the LCs were found to be sensitive to the relative humidity within the enclosure. Therefore, care should be taken to physically isolate the LCs from the accumulation of water vapor that may develop. The creation of a small aperture (5 cm by 30 cm) in the enclosure was useful for preventing such accumulation while maintaining environmental control.

To control the effect of equilibrium gas pressure[10] and potential effects of gas diffusivity[11] on acoustic cavitation, the quantity of dissolved gas was controlled via a degassing chamber (Onda Corp.). Dissolved oxygen (DO) content was monitored prior to and during studies using a DO meter (Oakton). The concentrations of other atmospheric gasses were assumed to be in consistent ratio with the quantity of dissolved oxygen. To allow for flexibility in gas composition in the synthetic ureter without affecting the acoustic path, an acoustic stand-off could be placed between the transducer and the synthetic ureter made of gelatin (~60 grams/L, Knox Gelatin). This acoustic stand-off was found to have a negligible effect on the transmitted pressure, as verified by needle hydrophone, mimicking the low-dissolved-gas intervening tissue present between the ureter and skin while permitting variable gas content in the synthetic ureter.

**F.  Automation**

For precise synchronization, a microcontroller (Teensy 3.2, Arduino) orchestrated the timing of syringe pump infusions, ultrasound output, and measurements made by LC1 and LC2, while a built-in function generator in the ultrasound energy source (TPO-102, Sonic Concepts) synchronized amplifier pulse synthesis. An RS232 interface with the syringe pump console as well as LC1 and LC2 was implemented by wiring the microcontroller to DB9 breakouts, from which DB9 cables were connected to the appropriate devices. The built-in function generator of the ultrasound energy source was connected with a simple transistor, allowing the microcontroller to trigger an ultrasound waveform via an interrupt function of the ultrasound generator.

**G.  Application of acoustic pressure**

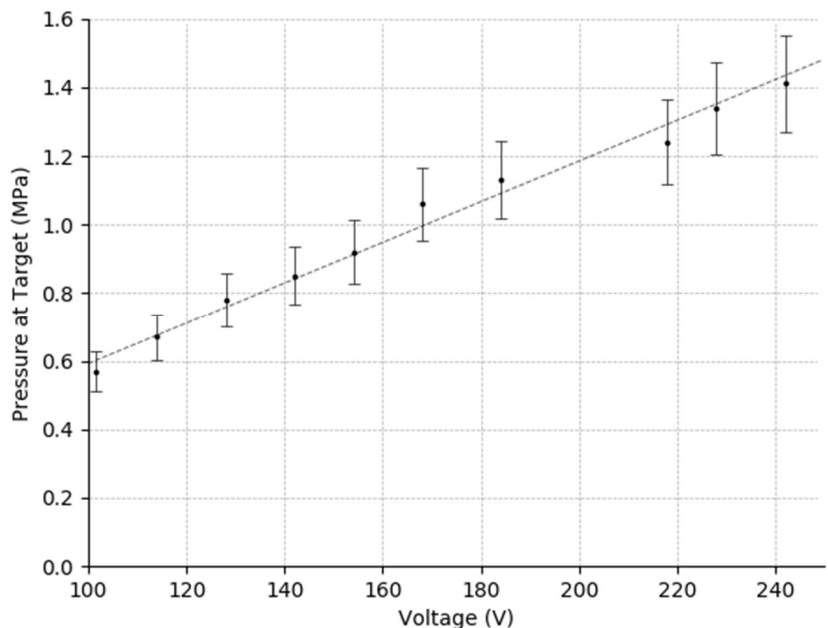

FIG. 3. The pressure produced at a point 10 cm from the face of the ultrasound transducer as a function of voltage across one of four identical piezoelectric elements shows the expected linear dependence with slope 5.9 kPa/V linear ($R^2 = 0.99$).

The pressure transducer used herein (XR-089, Sonic Concepts) is a research variant of a treatment device currently in clinical trials (Applaud Medical, Inc.). The transducer has a 1.25 cm aperture design that replicates the salient features of the clinical treatment transducer design, namely quasi-collimated distribution of acoustic energy, sub-MHz insonation frequency (400 to 600 kHz), and peak pressures up to 1.4 MPa. The broad pressure distribution of the transducer replicates another key feature of the clinical variant, namely ease of aiming without the aid of ultrasonic imaging techniques.

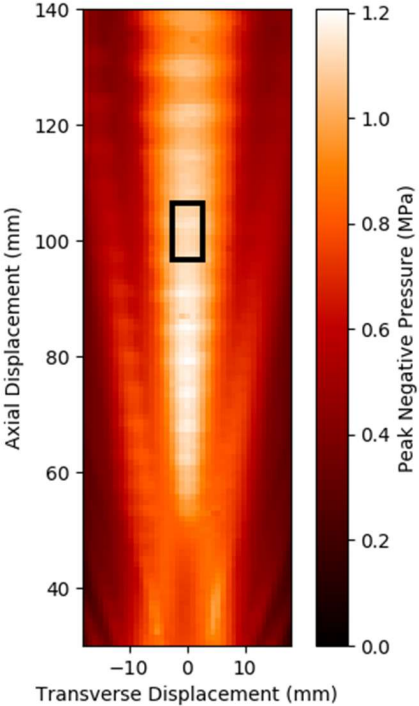

FIG. 4. The quasi-collimated beam produced by the acoustic energy source driven at 500 kHz was characterized. The center of the cylindrically symmetric energy source was located at the origin. The black outline represents the profile of a 5 mm by 10 mm stone positioned within the beam.

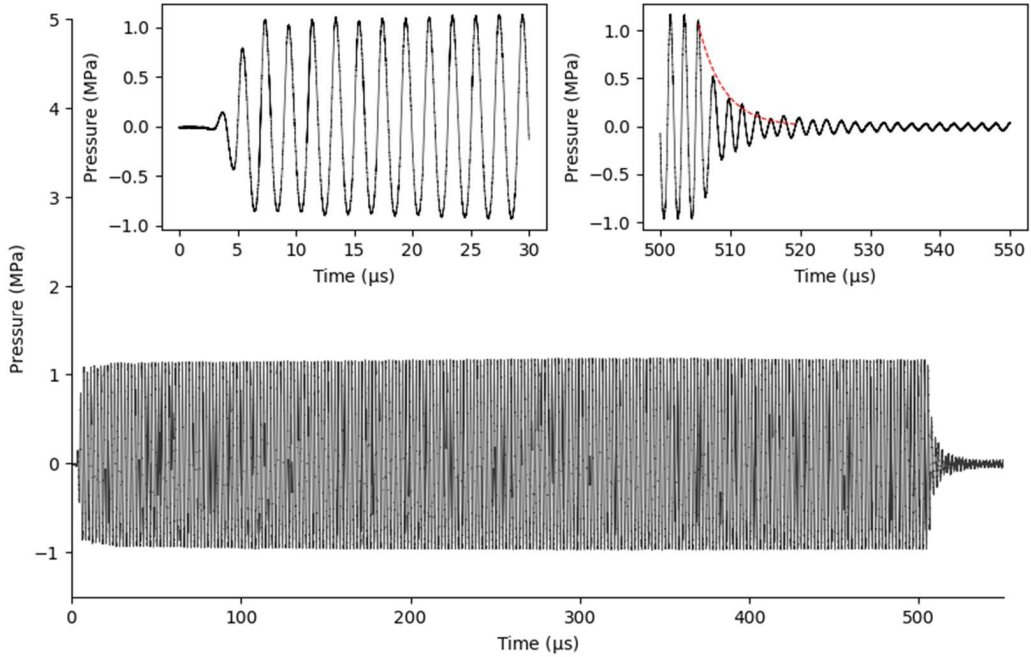

FIG. 5. Typical 500 kHz, 500 µs, approximately 1.1 MPa acoustic pulse recorded by a needle hydrophone (HNR, Onda Corp.) axially displaced 10 cm from the transducer. A typical pulse required two cycles to reach full amplitude. Q-factor of transducer was 5.7 as determined by fitting to an exponential ($R^2 = 0.97$).

The full spatial distribution of peak negative pressure generated by the acoustic energy source was characterized with millimeter resolution using three servo-controlled translation stages (T-LSM150A, Zaber Technologies) to position the calibrated HNR-0500 needle hydrophone in a large water tank lined with F28 absorbing material (Precision Acoustics) (Fig. 4). The measured pressure distribution at the point 10 cm from the face of the transducer exhibits axial uniformity, and the full-width, half-maximum (FWHM) of the beam in the transverse direction was found to be ~32 mm. Measured peak pressure was found to be linear with applied voltage (Figure 3). The maximum pressure at this point was 1.4 ± 0.1 MPa at 500 kHz, with a weak dependence on the axial dimension such that the pressure varied by 16.4% within a volume of a 5 mm diameter by 10 mm long cylindrical target.

A representative 500 µs acoustic pressure pulse is shown in Figure 5, with frequency 500 kHz and approximately 1.1 MPa pressure, measured at a position on-axis and 10 cm from the face of the transducer. A typical pulse required two cycles to reach full amplitude, and the Q-factor of transducer was measured to be 5.7 as determined by the "ring down" method by fitting to an exponential ($R^2 = 0.97$).

## III. RESULTS

### A. Bubble homogeneity

The effect of nutation on microbubble size distribution and concentration was studied. 10 mL samples of microbubbles at initial concentration 5e8 mL$^{-1}$ were nutated for 12 minutes. 1.0 mL aliquots were infused through a 5 Fr catheter (Open-End Ureteral Catheter, Cook Medical) at 5- and 12-minute time points and characterized using a multisizer (Beckman Coulter) with a 30-μm aperture. Each measurement was repeated in triplicate.

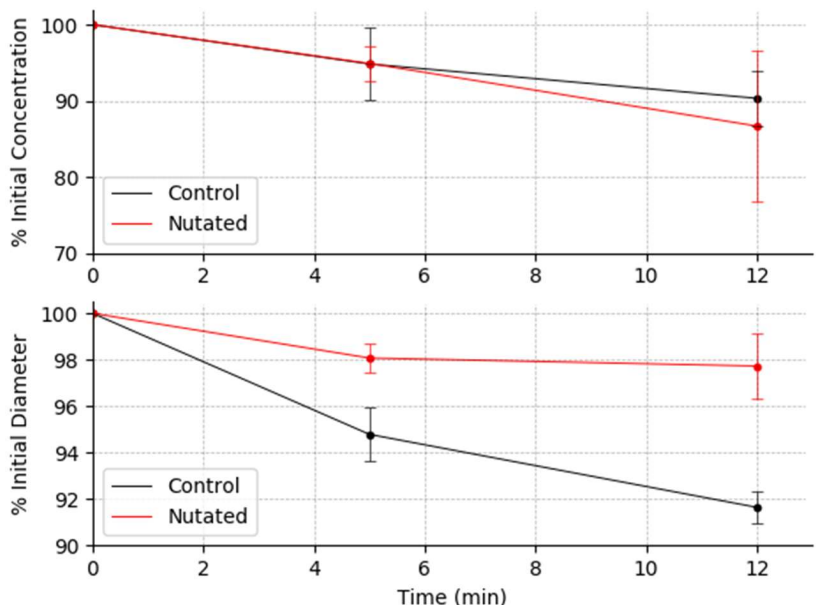

FIG. 6. Top: bubble size distribution is preserved with nutation with only marginal impact on total concentration relative to control. Bottom: the syringe pump fails to deliver larger bubbles if the sample is not nutated; this effect becomes more pronounced over time.

Multisizer (Beckman Coulter) characterization confirmed mean bubble diameter and concentration of samples within each test group were initially identical. Mean bubble diameter was 1.348 ± 0.005 μm and 1.335 ± 0.009 μm for control and nutated samples, respectively. Initial concentrations were 5.476 ± 0.185 e+6 mL$^{-1}$ and 5.336 ± 0.112 e+6 mL$^{-1}$, respectively. Fig. 4 illustrates the stability of bubble size distribution and concentration over 12-minute periods. The impact of nutation on concentration was not statistically significant relative to controls; however, nutation clearly preserved the mean bubble diameter of samples relative to controls, in which larger bubbles were disproportionately absent at the 5- and 12-minute timepoints.

## B. Characterization of instrument settling time

The stone-LC1 coupling was designed to displace minimally with the application of ultrasonic radiation force and achieve the shortest settling time to stable mass measurement. Accordingly, the calculi holder exposed a small profile to incoming acoustic waves, a feature that reduced damping due to hydrodynamic drag and minimized area that could be subjected to radiation forces.[12] Manufacturing the holder from wire was empirically found to reduce settling times.

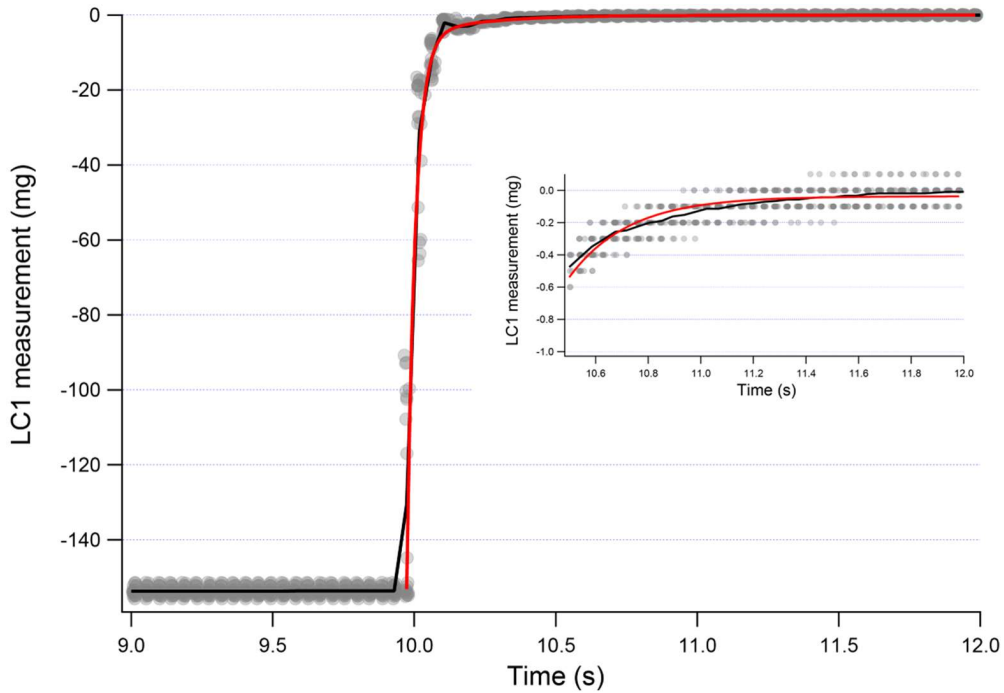

FIG. 7. Characterization of relaxation time of the coupled instrument-calculus system after insonation. Mass measurements (grey, N=21) were taken for 10 seconds during insonation and for the subsequent 5 seconds. The average settling time (black trace) was best fit by a double exponential (red) with time constant of 29.7 ± 0.6 ms for the initial phase and 226.5 ± 52.4 ms for the second phase. Note that mass readings on LC1 are negative during insonation due to the upwards radiation force.

The settling time of the instrument plus sample was characterized both to demonstrate that LC1 would return to zero after insonation and to determine the interval after which reliable measurement could be made. Toward this end, a cylindrical metal target of mass 1.554 g and surface area 46.2 mm² (facing the stone when suspended) served as a target for acoustic energy. Ultrasound was administered for 10 seconds ($f$=500 kHz, $p_A$=1.4 MPa, PD=500 μs, PRF=100 Hz). Mass was measured on a repeating interval of ~40 ms both during and after insonation to characterize settling time. In 21 trials, LC1 returned to ≤ 0.1 mg within 1.06 ± 0.20 s (mean ± STD) of the end of insonation. This decay was biphasic, with 96% of the amplitude in an initial phase with time constant of 29.7 ± 0.6 ms, followed by a slower phase, shown in Figure 7 inset, with a time constant of 226.5 ± 52.4 ms.

**C. Mass resolution of the instrument**

To demonstrate the resolution of the instrument, mass measurements were taken during application of ultrasound to a metal phantom of the same mass and surface area as previously mentioned. 1.0 mL phosphate buffered saline (PBS) was infused near the metal phantom over the course of 30 seconds at 2.0 mL/min via a Cook catheter to simulate bubble infusion. 1.0 mL of fluid was simultaneously withdrawn from elsewhere in the tank via Pump2 to maintain equal volume within the tank. In the subsequent 30 seconds, the target was treated with ultrasound ($f$=500 kHz, $p_A$=1.2 MPa, PD=100 μs, PRF=100 Hz). After 15 seconds, mass measurement was recorded.

Over 100 cycles of infusion and insonation, LC1 drifted linearly from zero under the influence of ultrasound at a rate of ~-1.3 mg/hour, a rate that was small compared to signals of interest. With the drift removed, the RMS noise was 0.1 mg.

### D. Controlling for evaporation/temperature

To characterize the resolution of the instrument, the noise spectrum was measured without ultrasound or infusions of engineered microbubbles. The spectrum was characterized at 30°C, monitored by a thermocouple (Omega), to illustrate the effects of dual-load-cell drift correction. Mass measurements were taken by LC1 and LC2 at 0.5 Hz for a period of 1.75 hours.

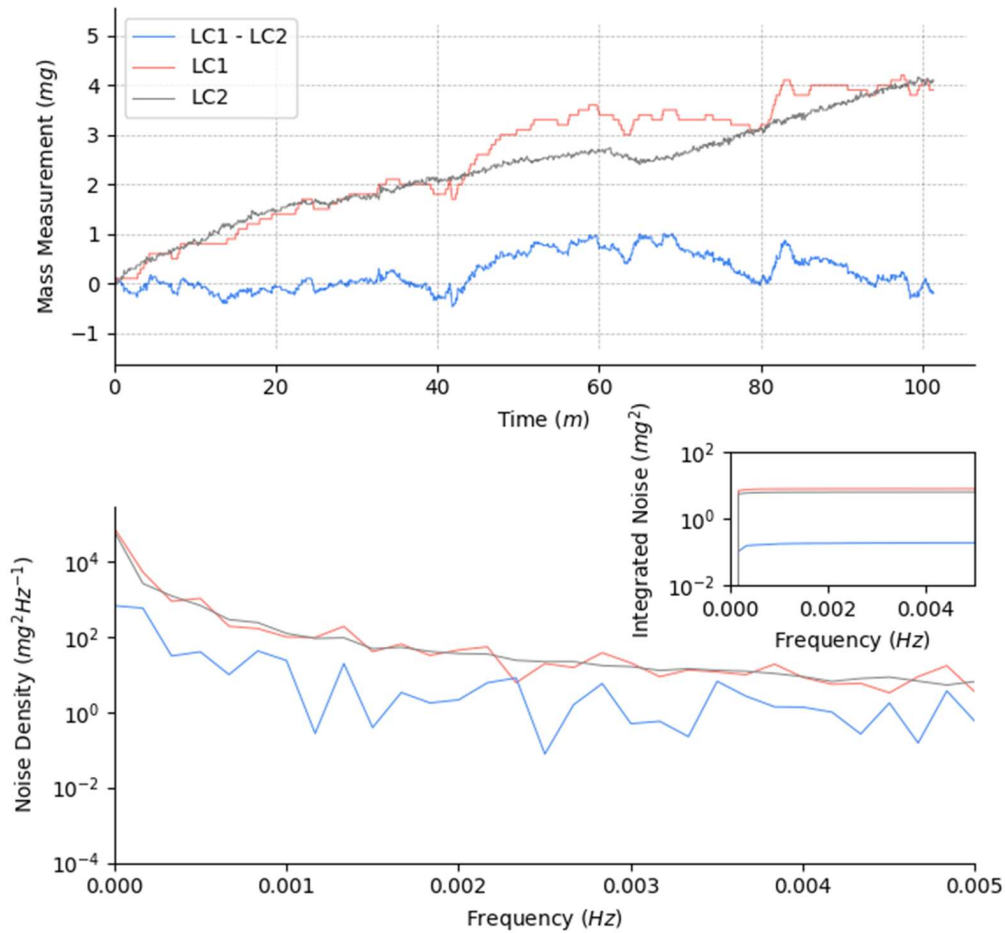

FIG. 8. Top: time-domain measurements of mass for LC1 (orange), LC2 (coupled to three underhook extensions, grey), and LC1 - LC2 (blue) at 30°C. Bottom: comparative noise power spectrums for LC1, LC2, and LC1 - LC2. Inset: integrated noise power for both measurement schemes. The RMS noise values for LC1, LC2, and LC1 - LC2 were 2.8, 2.5, and 0.4 mg, respectively.

The time-domain mass measurements, corresponding noise power spectra, and integrated noise densities for LC1, LC2, and LC1 - LC2 are shown in Fig. 8. LC2 is coupled to three carbon fiber rods that break the air-water interface, which resulted in a reduction in RMS noise compared to a single carbon fiber rod. Drift measured by LC1 and LC2 was observed to be temperature-dependent, consistent with evaporation-driven reduction of the buoyant force. From (8), the evaporation rate necessary to produce the observed drift was 5.5 μm/min; this value was comparable to the rate obtained from a pan test performed at equivalent temperature (5.1 μm/min). The square root of integrated noise for the differential measurement was ~0.3 mg as opposed to ~2.6 mg for LC1 alone.

**E. Validation of mass measurement**

### 1. Computing density of steel

To validate measurement of submerged mass via SCA, the masses of small zinc-coated steel pieces (~34 mg/piece) were measured with a separate analytical balance. The submerged mass of each piece was then measured by SCA. The two sets of measurements were linearly correlated with slope 1.15; from (6), the computed density of steel pieces was 7.82 g/ml, whereas the density of zinc-coated steel is reported as 7.86 g/ml.[13]

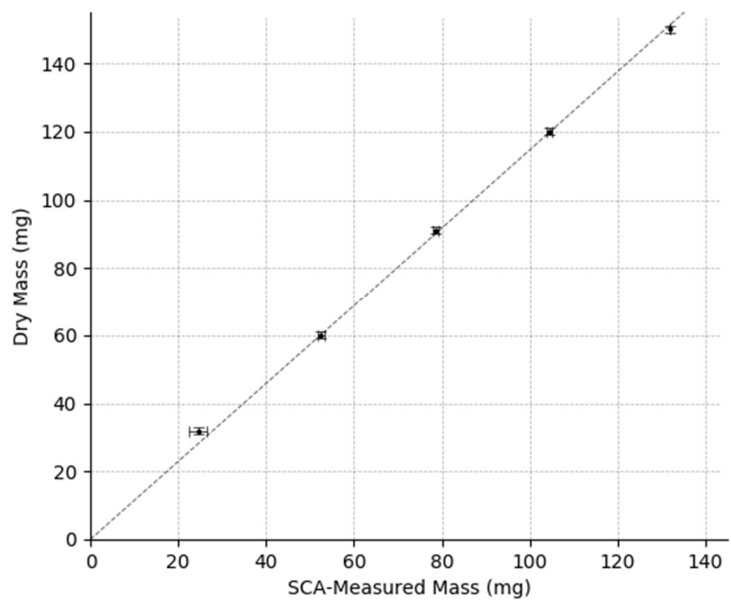

FIG. 9. Relationship between mass measurements of steel samples made by an analytical balance (ordinate) and SCA (abscissa). Linear regression of the data yields a slope of 1.15 ($R^2 = 0.998$), matching the coefficient predicted by (6), $\left(1 - \frac{\rho_L}{\rho_c}\right)^{-1}$.

### 2. Stone comminution

To further validate the presented method, the developed instrument was used to measure reductions in mass of 5 mm diameter model urinary stones (Riogen; 55% Hydroxyapatite, 20% $Al_2O_3$, 25% microcrystalline cellulose) treated with microbubble-mediated lithotripsy. Results were compared to differences in dry-weight measurements taken before and after experiments. Given (6), the two sets of measurements were expected to correlate linearly via a buoyancy correction factor, which was computed via linear regression of the data and by independent calculation using the average density of the population of calculi.

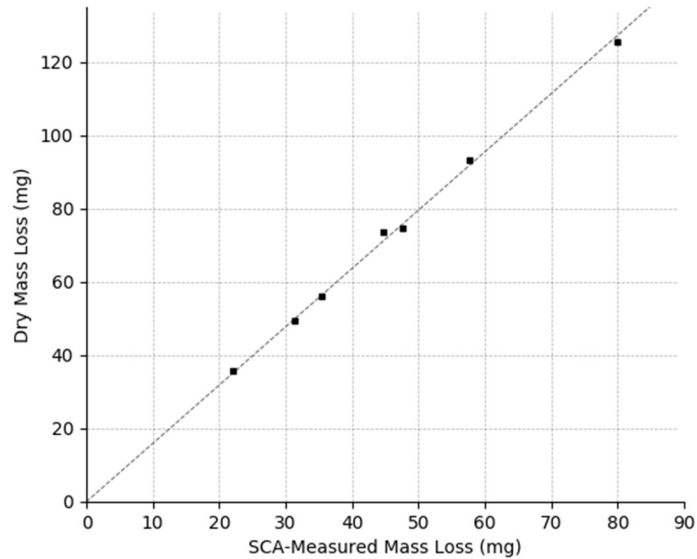

FIG. 10. Linear regression of measurements made by an analytical balance (of dried stones) and by SCA yielded a correlation constant of 1.59 ($R^2$ = 0.997), which matched the coefficient predicted by (6) where $\rho_c$ was estimated from the population of Riogen stones.

To measure average calculus density, samples were permitted to dry for 24 hours and subsequently weighed. Samples were then incubated in deionized (DI) water for 24 hours to remove trapped gas, which has been shown to induce artificial fragility in other phantoms.[5] The initial submerged weight of each calculus was then measured by placing it on the submerged holder. Calculi densities were computed via (7), and the average calculus density was 2.68 ± 0.28 (g/mL) (mean ± STD); the corresponding buoyancy correction factor was 1.60 ± 0.17.

To measure mass change effected by microbubble-mediated erosion, calculi were coupled to the instrument via a small rubber o-ring that gripped one extreme of a stone, allowing it hang and face the transducer. 0.5 mL infusions of microbubbles were delivered to each stone at 2.0 mL/min via Pump1 and catheter; 10-second doses of ultrasound ($f$=500 kHz, $p_A$=1.2 MPa, PD=500 μs, PRF=100 Hz) followed each infusion. Pump2 withdrew equivalent volumes of tank fluid to maintain equal volume within the tank. Mass changes measured by the instrument were recorded. Following treatment, stones were allowed to dry another 24 hours, after which the stones were weighed a final time.

Dry-weight measurements were plotted against corresponding submerged mass changes recorded by the instrument (Fig. 9), yielding a correlation constant of 1.59, which lay within the error of the buoyancy correction factor previously calculated. Average error in measurement was 1.8 ± 1.0 % (mean ± STD). Importantly, handling of stones may have increased the error in these measurements; the instrument may be more accurate than presented here.

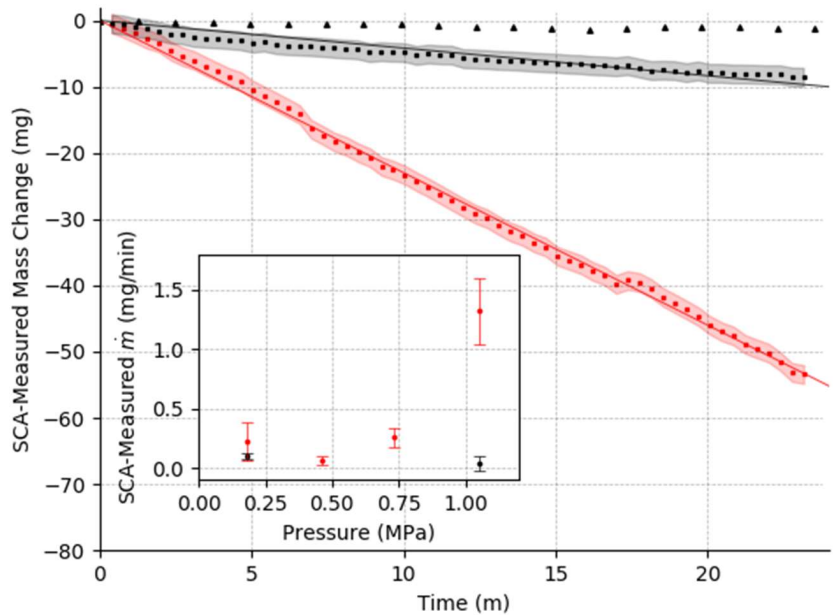

FIG. 11. The time-dependent SCA-measured mass change of calculi via lithotripsy with and without microbubbles (red and black, respectively). Black triangles represent drift while treating a non-eroding metal target. Engineered microbubbles were infused during a 15 second period, followed by acoustic insonation of 1.2 MPa applied in 500 μs pulses at a pulse repetition rate of 100 Hz for 5 seconds, followed by a 2 second measurement period. Inset: SCA-measured mass loss rate exhibits a strong dependence of driving pressure when microbubbles are present (red). Driving pressure without microbubbles does not produce calculus breakage.

To illustrate measurement of time-dependent mass loss mediated by calcium-adherent engineered microbubbles, microbubbles diluted to 5e8 mL$^{-1}$ were infused at a rate of 2.0 mL/min for 15 seconds. A 1.2 MPa, 500 kHz pulse train of 500 μs at 100 Hz PRF was used to insonate the microbubbles for a duration of 5 seconds. A 4 second delay between each microbubble infusion/insonation cycle was used to poll LC1 and LC2 in the absence of ultrasound or microbubble infusion before continuing on to the next cycle, resulting in the data shown in Figure 11. Fitting the microbubble-mediated mass change (Fig. 11, red) to a line produced a slope of $\dot{m}$ = -2.3X mg/min ($\sigma$ = 0.73 mg/min, $R^2$ = 0.998), which was 5.5-fold higher than the rate observed in the absence of microbubbles, -0.42 mg/min ($\sigma$ = 0.69 mg/min, $R^2$ = 0.877). There was an observed low-level drift of a non-eroding target ($\dot{m}$ = -0.04 mg/min, $\sigma$ = 0.21 mg/min, $R^2$ = 0.962), which had a negligible effect on the mass loss rate over this duration of measurement when reported to two significant figures.

To explore the dependence of $\dot{m}$ on applied pressure, transducer output pressure was varied under similar conditions (f=500 kHz, PD=500 μs, PRF=100 Hz with a cycle of 30 seconds of infusion followed by 30 seconds of insonation). The resulting erosion rate is shown in Figure 11, inset. A sudden increase in the mass loss rate was observed at ~1.1 MPa, consistent with work pursued by Maxwell *et al.,* in which stone breakage proceeded rapidly after a threshold driving pressure was met.[14] Importantly, this increase was not observed when bubbles are not present, indicating bubble-mediated pressure amplification lowers the threshold driving pressure necessary for breakage.

## IV. DISCUSSION

Quantifying time-dependent erosion is a difficult problem. In addition to the method presented here, the present authors considered two forms of image analysis—one of fallen stone fragments and the other of calculus debris collected via vacuum—as well as measurement of calculus mass by hand. While robust to environmental disturbances and capable of real-time measurement, image analysis presented the potential caveat of yielding results highly specific to a particular experimental setup. Previous work has shown that projected fragment area can be reliably measured,[5] but this approach requires human intervention to prevent sediment pile-up and is difficult to correlate with calculus mass because fragment size can vary widely. Measurement of stone mass by hand is laborious and requires removal of calculi from the experimental tank, which introduces ambiguous quantities of adherent water.

In contrast, quantifying breakage with SCA presents a number of immediate benefits. Measurement of a near-linear erosion rate when comminuting a homogenous calculus lends itself to statistical analysis, which can rapidly establish a high degree of confidence in the measured rate. Erosion rates may then be compared in parametric studies, and further, various parameters may be tested on the same stone. SCA also minimizes human labor and the handling of calculi, which can induce fallacious breakage and inadvertently change the relative orientation of transducer and calculus between trials.

Further improvements may extend SCA to shorter timescales. We hypothesize that the long, slow timescale for settling seen in Figure 7 is attributable to either the internal control system of the load cell, which produces a compensation current to return the load receiver to its zero position, or to the loose coupling between the calculus and holder provided by the cable. Specifically, the control system may be less than ideal for returning to equilibrium quickly when hydrodynamic drag is present. Future studies might shorten settling times with a purpose-built load-cell. Another feature of the current design is the versatility Pump2 accords in modeling physiological flow in the ureter, which impacts bubble localization to target calculi.

The overall design of this instrument has broader applications to submerged mass and force measurements. In the case of Radiation Force Balances (RFB) used to quantify acoustic output of ultrasonic equipment, the use of rigid carbon-fiber rods to hold a submerged target in a hanging orientation could be especially advantageous.

In conclusion, measurement of the mass of a submerged calculus via SCA is an accurate, high-throughput technique for quantifying stone erosion and fragmentation. Changes in submerged mass correlate well with changes in dry calculus mass provided subjects are of homogenous density. Here, we have illustrated the design for a device capable of intermittent insonation of a target material followed by measurements of buoyancy-reduced weight. These measurements are low-drift and accurate, and the method lends itself to high-throughput quantification of erosion rate for given input parameters. This technique has been applied to the quantitative measurement of microbubble-mediated erosion rate for an enhanced lithotripsy technique, illustrating the medical relevance of such benchtop experiments for the optimization of therapeutic treatments. Furthermore, the design principles illustrated here are more broadly applicable to underhanging mass and force measurements, such as typically encountered in Radiation Force Balance design. We anticipate that forthcoming results will investigate the relative importance of substrate mechanical strength, microbubble dynamics, and external driving parameters in facilitating microbubble-mediated comminution.